%% file: main.tex
\documentclass[fleqn,10pt]{wlscirep}
\usepackage[utf8]{inputenc}
\usepackage[T1]{fontenc}

\input{preamble.tex}
\DeclareMathAlphabet{\mathcal}{OMS}{cmsy}{b}{n}
\usepackage[singlelinecheck=false,justification=justified]{caption}

\graphicspath{ {fig/} }

\title{Quantum Machine Learning for Power System Stability Assessment}

\author[1]{Yifan Zhou}
\author[1,2,*]{Peng Zhang}
\affil[1]{Department of Electrical and Computer Engineering, Stony Brook University, NY 11794-2350, USA}
\affil[2]{Interdisciplinary Science Department, Brookhaven National Laboratory, Upton, NY 11973-5000, USA}

\affil[*]{p.zhang@stonybrook.edu}

\keywords{Quantum machine learning, Quantum classifier, Power system stability, Transient stability assessment.}

\begin{abstract}

Transient stability assessment (TSA), a cornerstone for resilient operations of today's interconnected power grids, is a grand challenge yet to be addressed since the genesis of electric power systems. 
This paper is a confluence of quantum computing, data science and machine learning to  potentially resolve the aforementioned challenge caused by high dimensionality, non-linearity and uncertainty. 
We devise a quantum TSA  (\QTSA{}) method, a low-depth, high expressibility quantum neural network, to enable scalable and efficient data-driven transient stability prediction for bulk power systems. 
\QTSA{} renders the intractable TSA straightforward and effortless in the Hilbert space, and provides rich information that enables unprecedentedly resilient and secure power system operations.
Extensive experiments on quantum simulators and real quantum computers verify the accuracy, noise-resilience, scalability and universality of \QTSA{}. \QTSA{} underpins a solid foundation of a quantum-enabled, ultra-resilient power grid which will benefit the people as well as various commercial and industrial sectors.

\end{abstract}
\begin{document}

\flushbottom
\maketitle

\thispagestyle{empty}

\section*{Introduction}

 Texas’ and California’s rolling outages~\cite{guardian2021,CA2020} in recent years signaled that our existing power infrastructures can hardly sustain the ever-expanding communities and deep integration of low-inertia renewables~\cite{doe2017quadrennial,heptonstall2020systematic}. 
The situations are rapidly deteriorating as our power grids are increasingly integrating massive DERs, such as intermittent rooftop solar photovoltaics (PVs), as well as solar farms and offshore wind systems, and have been subject to more frequent weather events~\cite{mills2016insurance,carreras2016north}.

A key technology to secure today's bulk power grids is transient stability assessment (TSA) which aims to determine the ability of the system to ride-through large disturbances (contingencies) and to reach the post-contingency steady-state~\cite{kundur1994power}. Transient instability is a fast phenomenon typically taking only a few seconds for the bulk system to collapse after contingencies occur. Due to this very nature, system operators in the control center never have sufficient time to steer the power system away from instability upon the occurrence of contingencies. For this reason, we have to rely on computer-based TSA without any manual interaction from human operators to assess the transient stability of the system~\cite{vaahedi2014practical}.

TSA, however, is a grand challenge yet to be addressed since the genesis of power systems in the era of Tesla, Edison and Westinghouse~\cite{kimbark1948power}. 
Interconnected power systems are the largest and most complicated man-made dynamical systems on this planet. Those bulk systems are highly nonlinear, exhibit multi-scale behaviors spatially and temporarily, and are increasingly stochastic and uncertain due to deep integration of renewable energy resources. The majority of the TSA methods being used in power industry rely on the explicit integration or implicit integration of differential equation models of the bulk power systems, which are known to be intractable to handle large power systems~\cite{iravani2020real,schafer2018dynamically,molnar2021asymmetry}. Discrete events such as frequent plug-and-plays of renewables, microgrids and loads in today's power systems~\cite{han2018stability,morstyn2018using} and unknown models due to data privacy issues~\cite{robu2019consider} make existing TSA methods even more computationally formidable even if they are executed on the powerful and expensive real-time simulators. 
Even worse, a large number of power system TSA must be conducted to examine the stability of the system in relation to massive `$N-k$' contingencies ($k$ components have failed in a power system with $N$ components), which further impedes the application of TSA in real-world power system operations.  
All the aforementioned challenges have made classical TSA prohibitively difficult for the online operation of large interconnected grids.

Today's power systems are undergoing an Enlightenment, where the confluence of big data, quantum computing and machine learning altogether is to drive a regime shift in the analysis and operation of our critical power infrastructures. Big data is the force behind the revolution: massive new types of intelligent electronic sensors such as synchronized phasor measurement units (PMUs), advanced metering infrastructure (AMI) meters and remote terminal units (RTUs)~\cite{phadke2008synchronized} are continuously generating gigantic volumes of data which allow for the development of data-driven power system analytics. Most recently, the successes in exploiting the potential of quantum supremacy~\cite{harrow2017quantum,arute2019quantum} shed lights on a `quantum leap' of computing capabilities. 
The power of quantum computing is derived from its ability to prepare and maintain complex superpositions of quantum states across many quantum degrees of freedom. While classically the number of required physical resources $N$ grows exponentially with the system complexity $n$, $N$ grows linearly with $n$ in a quantum computer, resulting in  exponential speedups over classical computing. 
Furthermore, highly entangled states, very difficult to represent on classical computers, are easily represented on a quantum computer~\cite{nielsen_chuang_2010,wiseman2009quantum}. Therefore, the intractable power system problems aforementioned, if formulated properly through programmable quantum circuits, can be executed efficiently on a quantum computer. 

Inheriting the exponential speedup of quantum computing in tensor manipulation~\cite{lloyd2013quantum},  the swift growth in quantum machine learning (QML) techniques~\cite{wittek2014quantum,biamonte2017quantum,schuld2018supervised} ignites new hopes
of developing unprecedentedly scalable and efficient data-driven power system analytics. 
QML is promisingly efficacious for data processing and model training in 
high-dimensional space that are intractable for classical algorithms~\cite{lloyd2014quantum,beer2020training}. 
Ideally, unique quantum operators such as superposition and entanglement, which can not be represented by classical operators, enable a superior representation of complicated data relationships~\cite{havlivcek2019supervised,schuld2021effect,lloyd2020quantum}. 
Nevertheless, the existing noisy quantum devices are still restrictive, hindering the implementing of QML if deep quantum circuits are needed.

This paper is the first attempt to unlock the potential of QML for power system TSA. 
A low-depth, high expressibility quantum neural network (QNN)-based transient stability assessment (\QTSA) method is devised  to enable scalable, reliable and efficient data-driven transient stability prediction.
In particular, we are focusing on designing an efficient \QTSA{} circuit that is feasible to pursue on near-term devices, considering the noisy-intermediate-scale quantum (NISQ) era~\cite{boixo2018characterizing,brooks2019beyond}, but general enough to be directly expandable to the noise-free quantum computer of a distant future (5-10 years). We have designed systematical studies which have demonstrated the robustness, accuracy and fidelity of \QTSA{} on real-scale power systems. Our \QTSA{} has shown consistently high performance on quantum computers at different noise levels.

Stability assessment plays a central role in nearly every field of the life sciences, physics and engineering. Our \QTSA{} therefore is promising to positively impact the modeling, analysis, controlling and securing various high-dimensional, nonlinear, hybrid dynamical systems. In particular, \QTSA{} underpins a solid foundation of a quantum-enabled resilient power grid, i.e., tomorrow's unprecedentedly autonomic power infrastructure towards self-configuration, self-healing, self-optimization and self-protection against grid changes, renewable power injections, faults, disastrous events and cyberattacks. Such a quantum-enabled grid will benefit the people as well as various commercial and industrial sectors.

\section*{Results}

Power system transients are generically modelled as a set of nonlinear differential algebraic equations:
\begin{subequations} \label{equ:power:DAE}
\begin{align}[left = \empheqlbrace\,]
& \Dot{{X}} = {F}_D({X},{Y}) \label{equ:power:DAE:1} \\
& {0} = {F}_A({X},{Y}) \label{equ:power:DAE:2}
\end{align}
\end{subequations}
where ${X}$ and  ${Y}$ separately denote the differential variables and algebraic variables; \eqref{equ:power:DAE:1} formulates the nonlinear dynamics of power devices, such as generators (e.g., synchronous machines, distributed energy resources), controllers (e.g., governors, exciters, inverters), power loads, etc;  \eqref{equ:power:DAE:2} formulates the instantaneous power flow of the entire power grid.

TSA appraises a power system's capability of resisting large disturbance~\cite{kundur1994power}. 
Denote ${Z} = ({X},{Y})$, and $\phi(t,{Z})$ as the orbit of \eqref{equ:power:DAE} starting from ${Z}$.  An asymptotically stable equilibrium point (SEP) ${Z}_s$ of \eqref{equ:power:DAE} satisfies that: (a) ${Z}_s$ is Lyapunov stable; (b) there exists an open neighborhood $\mathcal{O}$ of ${Z}_s$ such that $\forall {Z} \in \mathcal{O}$ converges to ${Z}_s$ when $t$ approaches infinity~\cite{chiang2015stability}.
The stability region of ${Z}_s$  encloses all the states that can be attracted by ${Z}_s$ within an infinite time: 
\begin{equation} \label{equ:power:stability region}
    \mathcal{A}({Z}_s) = \{{Z} \in \mathbb{R}^n: \lim_{t\rightarrow \infty} \phi(t,{Z}) = {Z}_s\}
\end{equation}

Stability region theory states that system stability after a large disturbance is determined by whether the post-disturbance state is within the stability region of an SEP~\cite{chiang2015stability}. Therefore, 
to formulate the data-driven TSA, the idea is to establish a direct mapping between the post-disturbance power system states and the stability results \cite{james2017intelligent,paul2019pmu}.

The keystone of \QTSA{}, different from classical machine learning techniques, is that the transient stability features in a Euclidean space ${Z} \in \mathcal{E}$ are embedded into quantum states in a Hilbert space $\ket{\psi} \in \mathcal{H}$ through a variational quantum circuit (VQC), which serves as a QNN to explicitly separate the stable and unstable samples.

\begin{figure*}[!ht]
\centering
\includegraphics[width=\columnwidth]{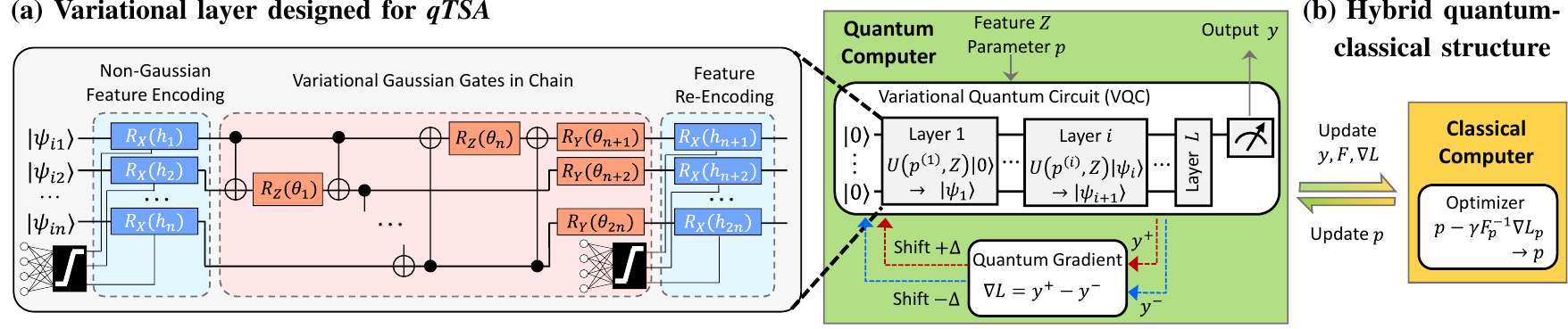}
\caption{\textbf{Design of low-depth, high expressibility quantum TSA.} 
\textbf{(a)}: Variational layer designed for \QTSA{}. Each layer $l$ consists of three blocks: a non-Gaussian feature encoding block, a variational  Gaussian entangled  block and a feature re-encoding block.
\textbf{(b)}: Hybrid training procedure of \QTSA{}. Quantum computer performs the feedforward execution of quantum circuits (i.e., measurement on the first qubit outputs the TSA result), and classical computer performs backward parameter update.}
\label{fig:main:variational algorithm}
\end{figure*}

Our key innovation is a design of \emph{low-depth, high expressibility} \QTSA{}, as presented in Fig.~\ref{fig:main:variational algorithm} (see also \textbf{Methods}). 
Fig.~\ref{fig:main:variational algorithm}(a) first visualizes the low-depth  variational \QTSA{} circuit which addresses the dimensionality and nonlinearity obstacles in TSA as well as the nonnegligible noise and source limitations on near-term quantum devices. Kernel ingredients in \QTSA{} circuit 
include (see \textbf{Methods}): 
\begin{enumerate}[label=\arabic*)]
\setlength\itemsep{-1pt}
    \item Non-Gaussian feature encoding: $\ket{\psi_{E}} = U_{E}({p}_{E}, {Z}) \ket{0}$ which adopts  parameterized, activation-enhanced quantum gates to enable a flexible, nonlinear and dimension-free encoding of power system stability features ${Z}$;
    \item Properly-arranged Gaussian quantum gates:  $\ket{\psi_{V}} = U_{V}({p}_{V})\ket{\psi_{E}}$ to efficiently represent the solution space;

    \item Feature re-encoding: $\ket{\psi} = U_{R}({p}_{R}, {Z})\ket{\psi_{V}}$ for enhanced expressibility of nonlinear behaviors in power systems;
    
    \item Repetitive layered structure: $\ket{\psi} = U(p,Z)\ket{0} \triangleq  U^{(L)}  \cdots U^{(2)} U^{(1)} \ket{0}$ to realize a more expressive and entangled VQC, where $U^{(l)} = U_{R}^{(l)} U_{V}^{(l)} U_{E}^{(l)}$ denotes the unitary operation at $l^{th}$ layer and $p$ assembles the circuit parameters at each layer.
\end{enumerate}

Then, as presented in Fig.~\ref{fig:main:variational algorithm}(b), \QTSA{} employs a hybrid quantum-classical framework for QNN training. The parameterized \QTSA{} circuit is executed on a quantum computer as the feedforward functionality of QNN, and parameter optimization is executed on a classical computer as the backpropagation  functionality. 
The two subroutines interact to train the VQC's parameters. Here, we introduce a generalized quantum natural gradient to enable a faster training of \QTSA{}: 
${p} - \dfrac{\eta}{\sqrt{\hat{v}_t \nabla\mathcal{L}_{{p}})} + \xi} \hat{m}_t {F}^{-1}_p \nabla\mathcal{L}_{{p}}  \rightarrow  {p}$, 
where $\mathcal{L}$ is the nonconvex cross-entropy loss representing the correctness of quantum-embedded power system stability results, $\nabla\mathcal{L}_{{p}}$ is the quantum gradient function and ${F}_p$ is the Fisher information matrix of the non-Gaussian quantum circuit (see \textbf{Methods}).

Extensive experiments of \QTSA{} on typical power systems ranging from benchmark grids to a real U.S. grid.  \QTSA{} is coded with Qiskit~\cite{Qiskit} and Pennylane~\cite{bergholm2018pennylane}, and is implemented on both a noise-free quantum simulator (\emph{ibmq\_qasm\_simulator}, a 32-qubit simulator) and a real IBM quantum device (\emph{ibmq\_boeblingen}, a 20-qubit machine).
Power system transient simulations with random fault locations and random fault clearing time are conducted by Power System Toolbox (PST) \cite{sauer2017power} to generate the stability features for \QTSA, such as power outputs, rotor angles and rotor speeds of generators, power flows through transmission lines,  and bus voltage angles and amplitudes.

\begin{figure*}[!t]
\centering
\includegraphics[width=\columnwidth]{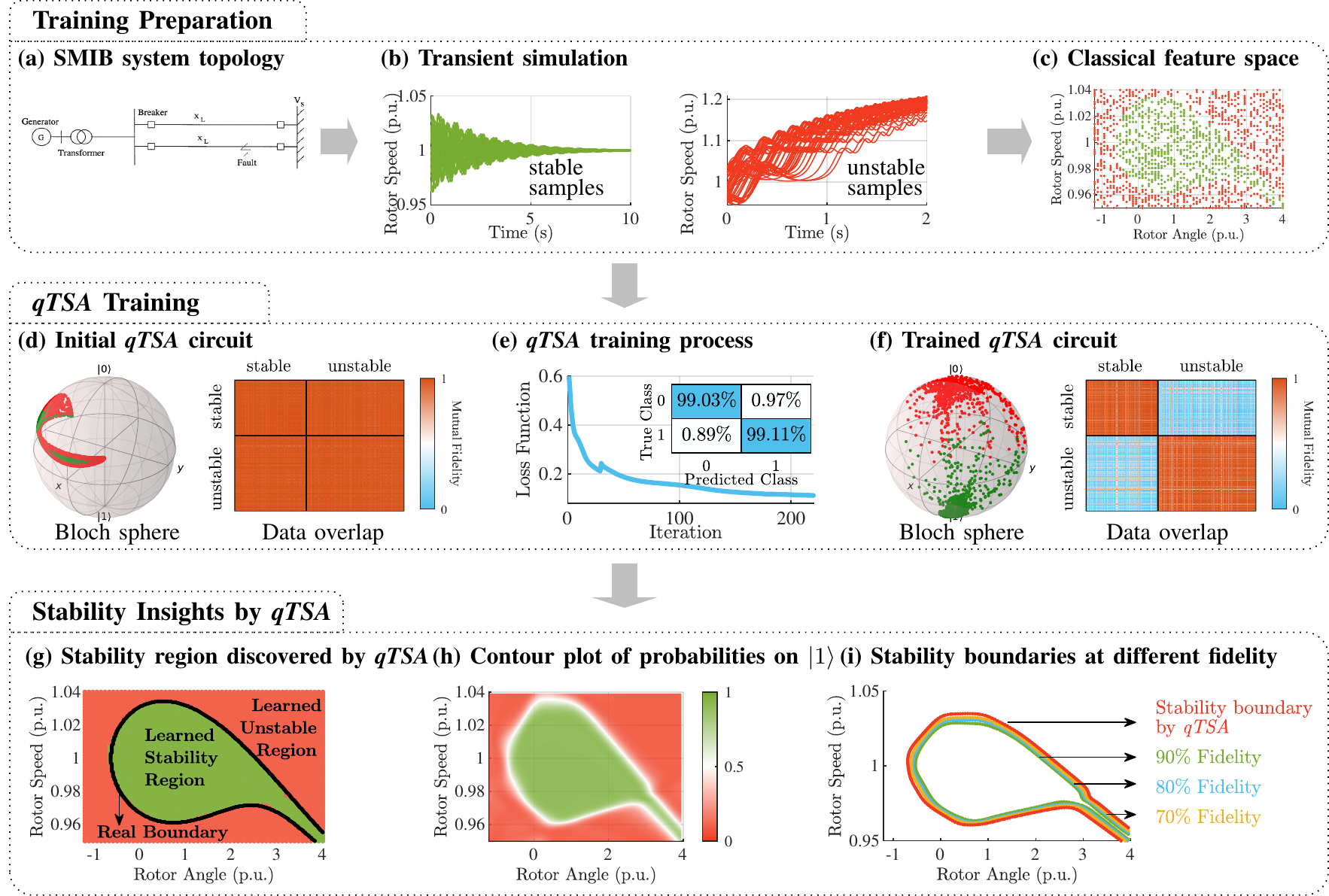}
\caption{\textbf{Demonstration of \QTSA{} procedure on the SMIB test system.}
Subplots (a)-(c) demonstrate training preparation.
Subplots (d)-(f) demonstrate \QTSA{} training.
Subplots (g)-(i) demonstrate the stability insights obtained by \QTSA{} learning.
\textbf{(a)}: Topology of the SMIB system.
\textbf{(b)}: Time-domain simulations of SMIB under random disturbances to generate stability features (i.e., rotor angle and rotor speed in this case).
\textbf{(c)}: Samples in the classical feature space (green for stable samples and red for unstable samples).
\textbf{(d)}: Samples embedded in the Hilbert space by a randomly-initialized \QTSA{} circuit. The quantum states are observed by both the Bloch sphere (i.e., a point $(\theta,\phi)$ on the Bloch sphere represents a one-qubit state $y=\alpha_y\bra{0}+\beta_y\bra{1}$ where $\alpha_y = cos(\theta/2)$ and $\beta_y=e^{i\phi}sin(\theta/2)$) and the data overlap (i.e., fidelity between quantum states corresponding to each pair of samples).
\textbf{(e)}: Loss function evolution during \QTSA{} training process and the final confusion matrix.
\textbf{(f)}: Samples embedded in the Hilbert space by the well-trained \QTSA{}, where the stable and unstable samples are distinctly separated. 
\textbf{(g)}: Stable and unstable regions discovered by the well-trained \QTSA{} and its comparison with the analytic stability boundary (i.e., black line).
\textbf{(h)}: Contour plot of probability on measuring $\ket{1}$ from the \QTSA{} circuit. Either a high probability on $\ket{1}$ or a high probability on $\ket{0}$ (i.e.,low probability on $\ket{1}$) reflects a high fidelity of transient stability prediction.
\textbf{(i)}: Stability boundaries discovered by \QTSA{} under different fidelity levels.
}
\label{fig:smib:QTSA}
\end{figure*}

The following sections investigate the capability of \QTSA{} in assessing power system stability on both noise-free simulators and near-term noisy quantum computers, as well as the performance of different quantum circuit designs for \QTSA.

\subsection*{Validity of \QTSA{} on Noise-Free Quantum Simulator} \label{sec:sim:qasm}

Three typical power systems are studied:
\vspace{-5pt}
\begin{itemize} 
\setlength\itemsep{-1pt}
    \item \emph{Single-machine infinite-bus (SMIB) system}, one of the most widely-used test systems in power system research. Since the stability region of SMIB (i.e., a 2-dimensional region) can be analytically computed, SMIB always serves as an indispensable benchmark for testing the performance of a TSA method.  
    \item \emph{Two-area system}, another most widely-used benchmark system exhibiting both local and inter-area oscillation modes. The system models and parameters are constructed from a real North American power system~\cite{kundur1994power}.
    \item \emph{Northeast Power Coordinating Council (NPCC) test system}, a real Northeastern US power system~\cite{sauer2017power} which was involved in the Northeast blackout of 2003.
\end{itemize}
\vspace{-5pt}

The overall \QTSA{}  procedure is exemplified via the SMIB system.  We then verify the efficacy of \QTSA{} on all the test systems.

\vspace{5pt}

\noindent \emph{\textbf{Training Preparation}}. Various dynamics samples are generated by performing transient simulations on the disturbed SMIB system (see Fig.~\ref{fig:smib:QTSA}(a)).  
Fig.~\ref{fig:smib:QTSA}(b) shows both the stable cases with dampened oscillations and the unstable cases where power generators run out-of-step. 
The initial states of the post-disturbance system, as shown in Fig.~\ref{fig:smib:QTSA}(c), constructs the feature set for \QTSA{} training. Identifying the stability region in the classical feature space is generally an intractable task.
Our hope is that \QTSA{} can directly distinguish the stable and unstable samples in the Hilbert space.

\vspace{5pt}

\noindent \emph{\textbf{\QTSA{} Training}}.
A \QTSA{} circuit with 2 qubits, 6 layers (denoted by \QTSA(2,6)) is constructed for the SMIB system. Fig.~\ref{fig:smib:QTSA}(d) to (f) show how an untrained  \QTSA{}  circuit evolves into a well-trained one.  We use Bloch sphere and quantum states overlap to evaluate the \QTSA{} efficacy in the Hilbert space.

Starting with a randomly-initialized \QTSA{} circuit, Fig.~\ref{fig:smib:QTSA}(d) shows that the stable and unstable samples are mixed on the Bloch sphere and quantum states corresponding to stable and unstable samples are highly overlapped, meaning the untrained circuit cannot be used for TSA.
Then, Fig.~\ref{fig:smib:QTSA}(e) shows the evolution of loss function during the \QTSA{} training, where the classification accuracy exceeds 99\% for both stable and unstable samples at the final stage.
The Hilbert space observations in Fig.~\ref{fig:smib:QTSA}(f) demonstrate the efficacy of the trained \QTSA. It can be seen that the well-trained quantum circuit successfully embeds the stable (unstable) samples to the lower-(upper-) half Bloch sphere. As can be seen from the data overlap, the mutual fidelity between samples from the same class is almost 1, while that from different classes is close to 0. Both observations verify a successful \QTSA{} being able to clearly separate the opposite classes through the VQC. 

\vspace{5pt}

\noindent \emph{\textbf{Stability Insights by \QTSA{}}}. 
A unique feature of \QTSA{} is that it offers not only the stability classification results but also the fidelity of stability or instability, by virtue of the multi-shot execution of VQC. 

First, \QTSA{} is able to accurately discover power system stability regions $\mathcal{A}(Z_s)$ (see \eqref{equ:power:stability region}).
As shown in Fig.~\ref{fig:smib:QTSA}(g), there is a faithful match between the stable/unstable regions learned by \QTSA{} (i.e., the green and red regions) and those obtained from analytical solutions.
Stability regions discovered by \QTSA{} allow power system operators to identify or predict the system stability under arbitrary contingencies in real-time. For instance, the SMIB system will remain stable if its post-contingency state (a vector of rotor angle and rotor speed in this case) locates in the green zone, whereas it must collapse if the post-contingency state locates in the red zone.

Second, the \QTSA{} results contain the probabilities of system stability. As shown in Fig.~\ref{fig:smib:QTSA}(h), for a post-contingency state falling in the stable zone, the farther it deviates from the stability boundary, the system stability ($\ket{1}$) would be assured with a higher probability. Similarly, the system would collapse ($\ket{0}$) with a higher probability if its state lies in the deeper red zone. The operating points with probabilities near 0.5 indicate marginal cases, which are observed to be located near the stability boundary. The probabilities associated with system stability provide valuable new information for system operators and decision makers to make more suitable planning, operation and remedial action schemes. Either the risk-averse or risk-takers will then selectively use such information to optimize the social welfare and improve electricity resilience.

Further, \QTSA{} is able to calculate stability regions at different risk levels.
Fig.~\ref{fig:smib:QTSA}(i) exhibits the stability boundaries corresponding to different fidelity thresholds. Being less risk-tolerant leads to shrunken stability regions because higher fidelity levels are adopted.

\begin{figure*}[!t]
\centering
\includegraphics[width=\columnwidth]{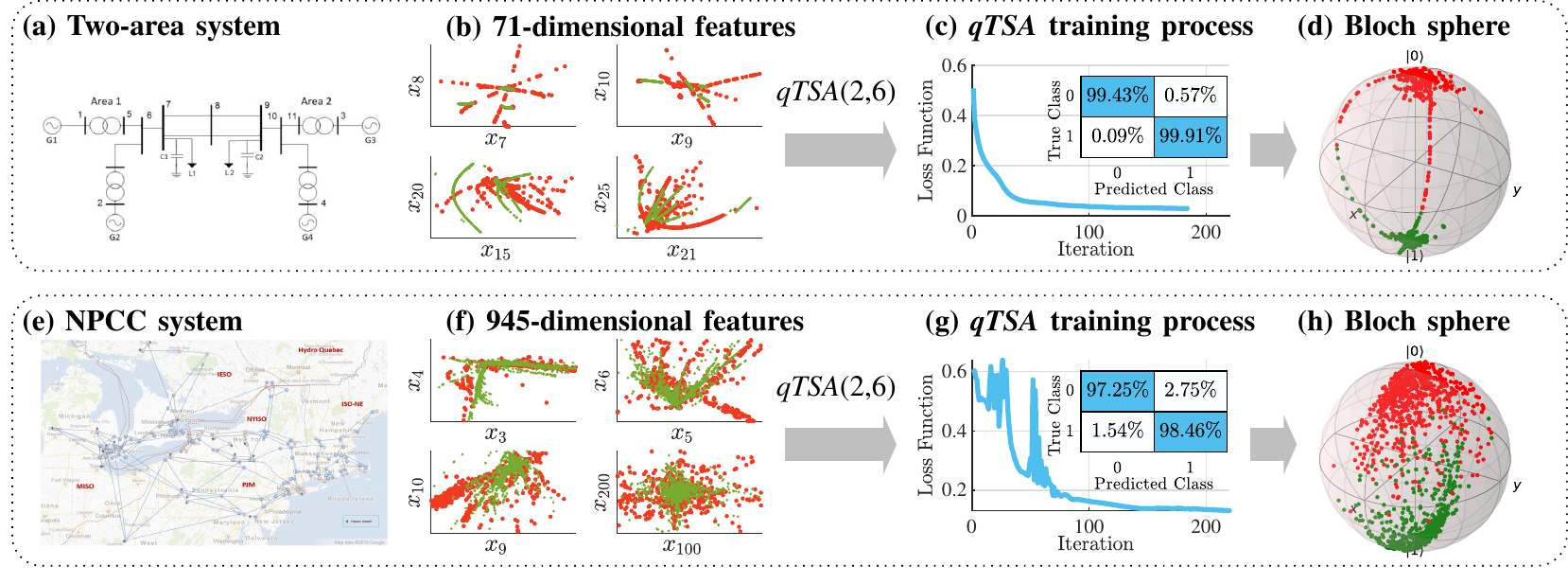}
\caption{\textbf{\QTSA{} results for practical power systems}.
Subplot (a)-(d) present \QTSA{} of the two-area system.
Subplot (e)-(h) present \QTSA{} of the NPCC system. Both cases use a \QTSA{}(2,6) circuit, where 2 qubits and 6 layers are employed.
\textbf{(a)(e)}: Power grid topology. 
\textbf{(b)(f)}: Samples in the classical space (green for stable samples and red for unstable samples). In both cases, stability features include power outputs, rotor angles and rotor speeds of generators, power flows through transmission lines,  and bus voltage angles and amplitudes.
\textbf{(c)(g)}: Loss function evolution during \QTSA{} training process.
\textbf{(d)(h)}: State measurements of the trained \QTSA{}. Stable samples (green) and unstable samples (red) are embedded to different halves of the Block sphere.
}
\label{fig:sim:QTSA}
\end{figure*}

\begin{table}[!t]
\caption{\textbf{Comparison of accuracy: \QTSA{} vs. TSA based on classical machine learning methods}}
    \label{tab:sim accuracy}
    \centering\small 
    \begin{threeparttable}
    \begin{tabular}{C{0.08\columnwidth}|C{0.12\columnwidth}|C{0.07\columnwidth}C{0.07\columnwidth}C{0.07\columnwidth}|C{0.07\columnwidth}C{0.07\columnwidth}C{0.07\columnwidth}}
    \hline \hline
    \multicolumn{2}{c|}{ }& \multicolumn{3}{c|}{ Training Set } & \multicolumn{3}{c}{ Test Set }  \\\cline{3-8}
    \multicolumn{2}{c|}{ }& \QTSA{}  &  SVM  & DNN & \QTSA{}  &  SVM  & DNN  \\\hline
     \multirow{4}{0.1\columnwidth}{\makecell{SMIB \\ System}} 
  &  Accuracy    &  0.9906  &  0.9525  &  0.9913  &  0.9810  &  0.9510  &  0.9865  \\ 
  &  Precision    &  0.9867  &  0.9672  &  0.9867  &  0.9726  &  0.9754  &  0.9855  \\ 
  &  Recall    &  0.9911  &  0.9185  &  0.9926  &  0.9820  &  0.9050  &  0.9820  \\ 
  &  F$_1$-Score    &  0.9889  &  0.9422  &  0.9897  &  0.9773  &  0.9389  &  0.9837  \\ 
     \hline
     \multirow{4}{0.1\columnwidth}{\makecell{Two-Area \\ System}} 
  &  Accuracy    &  0.9975  &  0.9931  &  0.9944  &  0.9860  &  0.9900  &  0.9850  \\ 
  &  Precision    &  0.9972  &  0.9917  &  0.9917  &  0.9795  &  0.9866  &  0.9779  \\ 
  &  Recall    &  0.9991  &  0.9981  &  1.0000  &  1.0000  &  0.9985  &  1.0000  \\ 
  &  F$_1$-Score    &  0.9981  &  0.9949  &  0.9958  &  0.9896  &  0.9925  &  0.9888  \\ 
     \hline
     \multirow{4}{0.1\columnwidth}{\makecell{NPCC \\ System}} 
  &  Accuracy    &  0.9800  &  0.9550  &  0.9765  &  0.9530  &  0.9470  &  0.9560  \\ 
  &  Precision    &  0.9831  &  0.9442  &  0.9822  &  0.9560  &  0.9455  &  0.9582  \\ 
  &  Recall    &  0.9846  &  0.9854  &  0.9798  &  0.9722  &  0.9757  &  0.9757  \\ 
  &  F$_1$-Score    &  0.9838  &  0.9644  &  0.9810  &  0.9640  &  0.9604  &  0.9669  \\ 
    \hline \hline
    \end{tabular}
\begin{tablenotes} \footnotesize
\item[1] $ \text{Accuracy} = \frac{\text{TP} + \text{TN}}{\text{TP} + \text{TN} + \text{FP} + \text{FN}}$; 
$ \text{Precision} = \frac{\text{TP} }{\text{TP} + \text{FP} }$; 
$ \text{Recall} = \frac{\text{TP} }{\text{TP} + \text{FN}}$; 
F$_1$-Score$ = 2 (\text{Precision}^{-1} + \text{Recall}^{-1} )^{-1}$. Here, TP, TN, FP, FN respectively denote the number of true positive, true negative, false positive and false negative samples.
\end{tablenotes}
\end{threeparttable}
\end{table}

\vspace{5pt}
\noindent
\emph{\textbf{Verification on Practical Power Systems}}.
We further exhibit the versatility and efficacy of \QTSA{} in the stability assessment of large power systems. Fig.~\ref{fig:sim:QTSA} presents the \QTSA{} results for the two-area system and the NPCC system. For both systems, as shown in Fig.~\ref{fig:sim:QTSA}(b) and Fig.~\ref{fig:sim:QTSA}(f), their high-dimensional features are irregularly distributed in the classical space and their stable/unstable samples are highly mixed. Once a trained \QTSA{} embeds the classical features into quantum states, surprisingly, it successfully arranges the stable and unstable samples onto the upper- and lower- halves of a Bloch sphere, rendering the challenging power system stability identification straightforward and effortless in the Hilbert space.

Table.~\ref{tab:sim accuracy} quantitatively evaluates the accuracy of \QTSA. For the small- and medium-scale power system, \QTSA{} achieves high accuracy on both the training set ($>99\%$) and test set ($>98\%$). 
Even for the large-scale NPCC system, \QTSA{} exhibits 
outstanding performance of 98\% accuracy on the training set and 95\% accuracy on the testing set.
\QTSA{} is further compared with two data-driven TSA methods based on classical machine learning algorithms, i.e., a support vector machine (SVM) with the radial basis function kernel, and a deep neural network (DNN) with a three-layer perceptron architecture. Both SVM- and DNN- based approaches are implemented in Scikit-learn~\cite{scikit-learn}. 
Both the $F_1$-scores and accuracy validate \QTSA{}'s competency and high performance in comparison with classical-machine-learning-empowered TSA approaches.

\subsection*{Comparison of Quantum Circuits}
\begin{figure*}[!t]
\centering 
\includegraphics[width=\columnwidth]{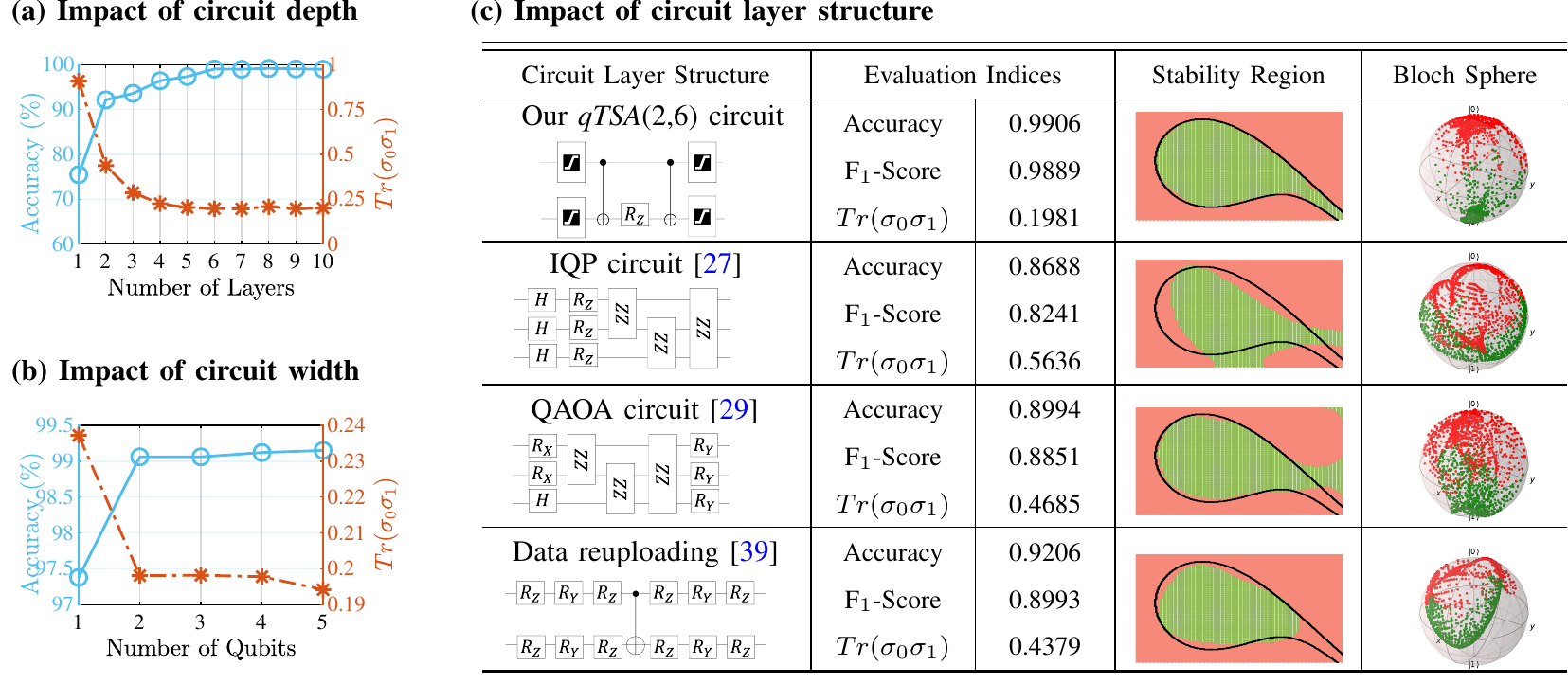}
\caption{\textbf{
\QTSA{} performances with different quantum circuit designs: A comparison}. Three indices are employed to evaluate the transient stability prediction capability of quantum circuits: classification accuracy, F$_1$-Score and $Tr(\sigma_0\sigma_1)$ (i.e., $\sigma_0 = \text{mean}_{i\in\mathcal{S}_0}(\ket{\psi_i}\bra{\psi_i})$ and $\sigma_1 = \text{mean}_{i\in\mathcal{S}_1}(\ket{\psi_i}\bra{\psi_i})$).
An improved separation between the stable and unstable classes is  indicated by improved accuracy, increased  F$_1$-Score and decreased $Tr(\sigma_0\sigma_1)$.
\textbf{(a)}, \textbf{(b)}: Performance of \QTSA{} circuits (\QTSA{} design architecture is illustrated in Fig.~\ref{fig:main:variational algorithm}) with different number of layers or qubits. 
\textbf{(c)}:  Performance comparison between our \QTSA{}(2,6) circuit and other typical QNN structures.  The IQP and QAOA circuits both employ 3 qubits and 10 layers. The data reuploading circuit employs 2 qubit and 10 layers. 
}
\label{fig:sim:IF circuit}
\end{figure*}

The realization of a quantized  TSA is by no means unique; rather, numerous quantum circuits could be devised for this purpose.  
In this section, we demonstrate how different designs of quantum circuits impact \QTSA's performance. Three factors are taken into consideration: circuit depth (i.e., number of layers), circuit width (i.e., number of qubits) and circuit layer structure.  In addition to classification accuracy and F$_1$-Score, another performance index $Tr(\sigma_0\sigma_1)$ is introduced for~\QTSA, where $\sigma_0 = \text{mean}_{i\in\mathcal{S}_0}(\ket{\psi_i}\bra{\psi_i})$ and $\sigma_1 = \text{mean}_{i\in\mathcal{S}_1}(\ket{\psi_i}\bra{\psi_i})$ respectively denote the mean density matrices of unstable/stable samples. A decrease in $Tr(\sigma_0\sigma_1)$ thus indicates an improved separation between the stable and unstable classes.
Without loss of generality, the comparisonal studies are performed on the SMIB system, as presented in Fig.~\ref{fig:sim:IF circuit}.

Subplots~\ref{fig:sim:IF circuit}(a) and~\ref{fig:sim:IF circuit}(b) jointly show the impacts of scaling factors such as circuit's depths and widths. 
The following insights can be obtained: 

\begin{itemize}
\setlength\itemsep{-1pt}
    \item Initially, increasing the number of layers tends to cause improved accuracy and better separations between stable/unstable samples.  
    The rationale behind this is an enhanced expressibility of the quantum circuit. Note that the performance of \QTSA{} starts to saturate once the depth goes beyond a certain number (6 in this case). 
    \item Increasing the number of qubits also leads to improved \QTSA{} performance, whereas the expressibility saturation is again observed. A noteworthy observation is that \QTSA{} 
    manifests impressive expressibility even with a single qubit, where the classification accuracy still reaches 97.38\%.
    \item Although more layers and qubits definitely 
    boost the expressibility of \QTSA, an overscale quantum circuit is not recommended.
    Such a circuit would demand prohibitively expensive quantum resources due to the saturation phenomenon.
    It also aggravates the burden of QNN training accompanied with the increased parameters of the variational circuit. 
    Finally, a large-scale quantum circuit is prone to the noisy quantum environment,  leading to a sharply declined TSA performance on real quantum devices (which will be further discussed in the next subsection).

\end{itemize}

Then, Fig.~\ref{fig:sim:IF circuit}(c) compares the performance of our  low-depth, high expressibility \QTSA~circuit with those of three typical QNN structures, i.e., an Instantaneous Quantum Polynomial (IQP) circuit \cite{havlivcek2019supervised}, the Quantum Approximate Optimization Algorithm (QAOA) inspired circuit \cite{lloyd2020quantum}, and a data reuploading circuit \cite{perez2020data}.  
Our \QTSA{} circuit exhibits the best expressibility benefiting from its non-Gaussian feature encoding and entanglement layers.

\subsection*{\textbf{Validity of \QTSA{} on Noisy, Real Quantum Computing Environment}}\label{sec:sim:ibmq}

\begin{table}[]
\caption{\textbf{Comparison of \QTSA{} accuracy: noisy real quantum device \emph{ibmq\_boelingen}  vs. noise-free simulator \emph{ibmq\_qasm\_simulator} }}
    \label{tab:ibmq}
    \centering\small 
    \begin{tabular}{C{0.11\columnwidth}|C{0.12\columnwidth}|C{0.1\columnwidth}C{0.15\columnwidth}|C{0.1\columnwidth}C{0.15\columnwidth}}
    \hline \hline
    \multicolumn{2}{c|}{ }& \multicolumn{2}{c|}{ Training Set } & \multicolumn{2}{c}{ Test Set }  \\\cline{3-6}
    \multicolumn{2}{c|}{ }& \emph{ibmq\_boelingen}  & \emph{ibmq\_qasm\_simulator} & \emph{ibmq\_boelingen}  & \emph{ibmq\_qasm\_simulator} \\\hline
     \multirow{4}{0.11\columnwidth}{\makecell{SMIB \\ System}} 
  &  Accuracy    &  0.9819  &  0.9906  &  0.9770  &  0.9810  \\ 
  &  Precision    &  0.9708  &  0.9867  &  0.9679  &  0.9726  \\ 
  &  Recall    &  0.9867  &  0.9911  &  0.9772  &  0.9820  \\ 
  &  F$_1$-Score    &  0.9787  &  0.9889  &  0.9725  &  0.9773  \\ 
     \hline
     \multirow{4}{0.11\columnwidth}{\makecell{Two-Area \\ System}} 
  &  Accuracy    &  0.9907  &  0.9975  &  0.9910  &  0.9930  \\ 
  &  Precision    &  0.9882  &  0.9972  &  0.9866  &  0.9896  \\ 
  &  Recall    &  0.9980  &  0.9991  &  1.0000  &  1.0000  \\ 
  &  F$_1$-Score    &  0.9931  &  0.9981  &  0.9933  &  0.9948  \\ 
     \hline
     \multirow{4}{0.11\columnwidth}{\makecell{NPCC \\ System}} 
  &  Accuracy    &  0.9710  &  0.9800  &  0.9450  &  0.9540  \\ 
  &  Precision    &  0.9727  &  0.9831  &  0.9414  &  0.9527  \\ 
  &  Recall    &  0.9806  &  0.9846  &  0.9772  &  0.9787  \\ 
  &  F$_1$-Score    &  0.9767  &  0.9838  &  0.9590  &  0.9655  \\ 
    \hline \hline
    \end{tabular}
\end{table}

\begin{figure*}[!t]
\centering 
\includegraphics[width=\columnwidth]{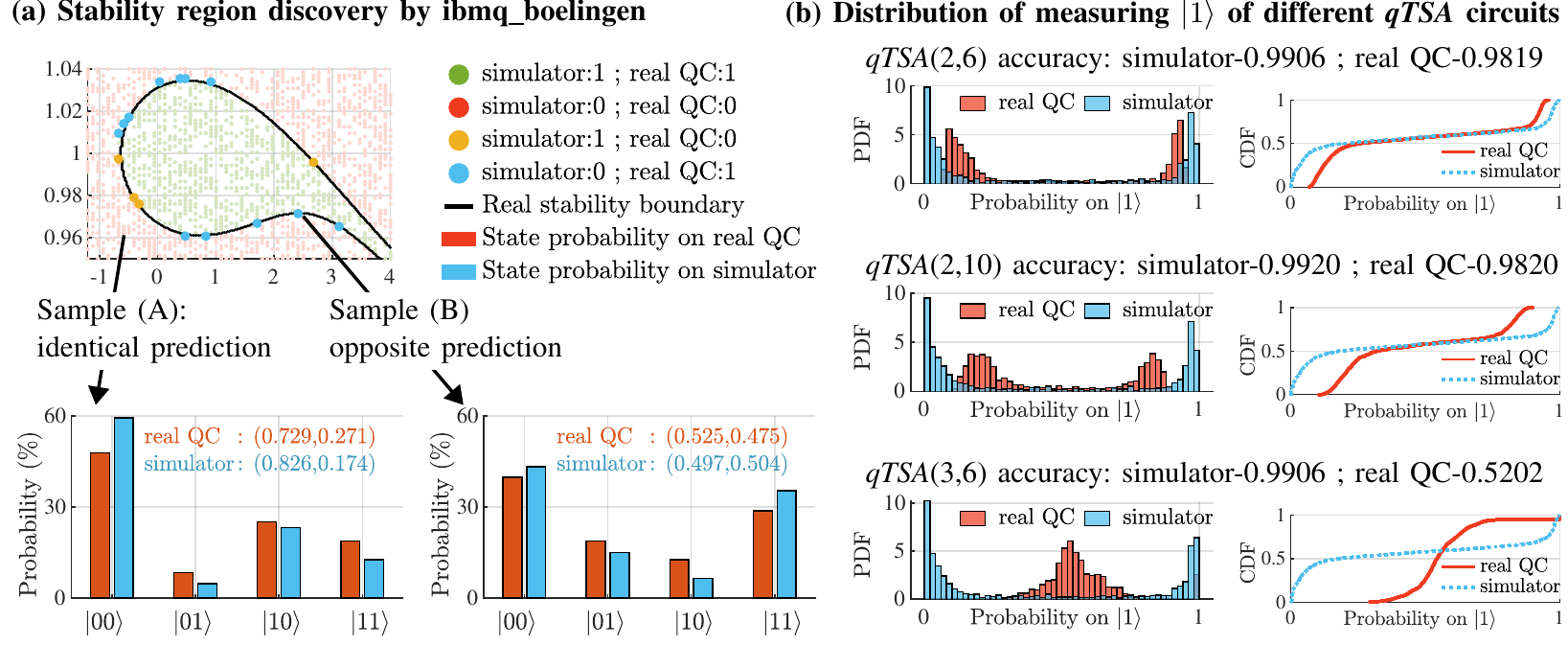}
\caption{\textbf{Implementation of \QTSA{} on the noisy IBM Quantum computer (\emph{ibmq\_boelingen})}. 
\textbf{(a)}:  Stability predictions obtained on the simulator \emph{ibmq\_qasm\_simulator} and the real noisy QC \emph{ibmq\_boelingen}. Accordingly, there are four types of results: green dots (samples predicted as stable on both devices); red dots (samples predicted as unstable on both devices); yellow dots (samples predicted as stable on the simulator but unstable on the real QC); blue dots (samples predicted as unstable on the simulator but stable on the real QC). 
Two typical samples exemplify the quantum state probability measured from \QTSA{} circuit. Sample (A) illustrates an identical prediction on both devices, which represents the most cases (green and red dots), whereas sample (B) illustrates an opposite prediction, which is observed mainly around the stability boundary. 
\textbf{(b)}: Probability distribution function (PDF) and cumulative distribution function (CDF) of the probability of measuring $\bra{1}$ from different-scaled \QTSA{} circuits.
}
\label{fig:sim:real QC}
\end{figure*}

Today's quantum devices are limited by gate errors \cite{cross2019validating}, which hinders the implementation of deep quantum circuits. As an ultimate test of the practicality of \QTSA{}, we now implement it on a real quantum computer and systematically verify it under noisy quantum computing environments.  

We run \QTSA{} on the IBM Quantum device \emph{ibmq\_boelingen} which is a 20-qubit superconducting quantum computer (QC). 
Table.~\ref{tab:ibmq} presents \QTSA's accuracy obtained from \emph{ibmq\_boelingen}.  Compared with those results on the noise-free IBM Quantum simulator \emph{ibmq\_qasm\_simulator}, a surprising  finding is that a noisy real quantum computing environment has little effect on the performance of \QTSA{}  
and the overall classification is still of high quality. Even in the most challenging case of the real NPCC system,  only a 1\% decrease in the accuracy is reported, and the \QTSA{} results remain compelling as compared with those from classical machine learning algorithms. This experiment exhibits the effectiveness of \QTSA{} on the near-term  quantum devices and verifies the inherent resilience of 
\QTSA{} against noisy quantum computing environments.

Fig.~\ref{fig:sim:real QC}(a) further investigates the rationale behind the resilience of \QTSA, where the stability identification results are again obtained by \QTSA(2,6). It can be seen that samples of inconsistent prediction from \emph{ibmq\_boelingen} and \emph{ibmq\_qasm\_simulator} (highlighted by the blue and yellow dots) are mostly distributed around the stability boundary. 
The measurement probability distribution of two typical samples illustrated in Fig.~\ref{fig:sim:real QC}(a) shows that the measurement probabilities from \emph{ibmq\_boelingen} and \emph{ibmq\_qasm\_simulator} do not differ much. This indicates \QTSA(2,6) is only slightly perturbed by the noisy quantum device. Therefore, only for the low-fidelity areas where the quantum circuit output  already has a similar probability on both $\ket{0}$ and $\ket{1}$, a slight perturbation 
from the noisy environment would possibly produce an opposite classification result, as exemplified by sample (B). According to our previous study in Fig.~\ref{fig:smib:QTSA}(h), those samples locate in a narrow area around the stability boundary, which are actually of low stability margin, and thus their stability by their very nature 
is uncertain. 
Whereas, for the high-fidelity areas as exemplified by sample (A), the quantum circuit output in the noisy device  does not change the final \QTSA{} prediction. Therefore, \QTSA{} generate reliable results for most samples even on noisy QCs.

Even though $\QTSA(2,6)$ maintains high accuracy, a larger  scale \QTSA{} circuit may fail to produce similar results. Fig.~\ref{fig:sim:real QC}(b) illustrates the performance of \QTSA{}  circuits of different scales  on noisy QCs, where \QTSA(2,6) as the default case, \QTSA(2,10) with 10 layers and \QTSA(3,6) with an additional qubit are compared. 
\begin{itemize}
\setlength\itemsep{-1pt}
    \item  On the noise-free \emph{ibmq\_qasm\_simulator}, all circuits achieve high accuracy over 99\% with a notable feature that the quantum measurements mainly accumulate around 0 and 1. This indicates that most samples are predicted at high confidences, which again is coincident with  Fig.~\ref{fig:smib:QTSA}(h), i.e., only the narrow areas around the stability boundary are of low fidelity.
    
    \item For \QTSA(2,6) running on the noisy \emph{ibmq\_boelingen}, the peaks of the probability distribution  slightly shift towards the center, indicating a slightly decreased prediction fidelity. 

    \item For \QTSA(2,10), although its accuracy on the noisy QC remains comparable to that of \QTSA(2,6), the shift of probability distribution  peaks is more obvious, reflecting a further decreased fidelity on the noisy prediction.

    \item For \QTSA(3,6) involving more qubits and accordingly deeper circuit depth, the accuracy on the real QC sharply deteriorates down to 52.02\%, making the evaluation on the noise-free simulator meaningless. The probability distribution  histogram shows most of the measurements
    center around 0.5, indicating that output states from \QTSA{} are nearly random due to the noisy environment and quantum decoherence.
    
    \item Consequently, a high quality \QTSA{} need to possess both high accuracy and  high fidelity. This is important to ensure a high tolerance against noises.
\end{itemize}

\begin{figure*}[!t]
\centering 
\includegraphics[width=\columnwidth]{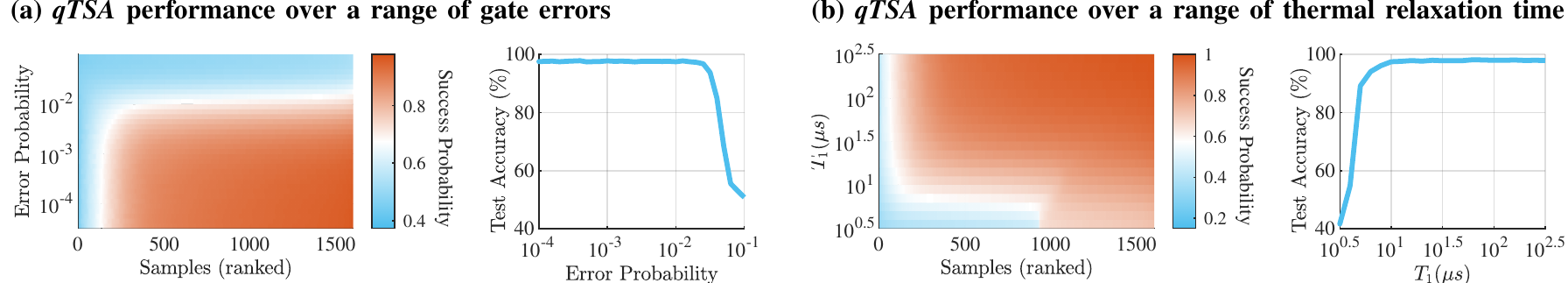}
\caption{\textbf{Impact of noisy quantum environment on \QTSA{} performance}. 
\textbf{(a)}: Left subplot shows \QTSA{}'s success probabilities of all samples (sorted from the sample of the lowest success probability to that of the highest) over a range of gate errors.  Right subplot shows \QTSA{}'s overall accuracy on the test set over a range of gate errors.
\textbf{(b)}: \QTSA{}'s success probabilities of all sample as well as the overall accuracy over a range of thermal relaxation time $T_1$ (i.e., the time it takes for a qubit to decay from the excited state to the ground state).
}
\label{fig:sim:error}
\end{figure*}

Since different quantum devices may have different noise features, it is necessary to systematically examine the \QTSA's performance under various noisy environments. 
We adjust the noise levels through a noise module provisioned by IBM and the noise settings are modified from the \emph{ibmq\_boelingen}'s real noise data. 
As Fig.~\ref{fig:sim:error} shows, excessive gate error and decohenrece time certainly lead to the collapse of the \QTSA{} circuit.
Nevertheless, high TSA accuracy can still be obtained with gate errors smaller than 0.02 or the thermal relaxation time smaller than $10\mu$s, which  can be easily satisfied by today's quantum hardware. The \QTSA{}'s excellent performance is therefore universal.

\section*{Conclusion}
This paper unlocks the potential of quantum machine learning in revolutionizing power system data-driven analytics. The
key innovation is to develop quantum neural network-based transient stability assessment to enable an ultra-scalable and efficient transient
stability analysis for resilient and secured decision-making of large-scale power systems. The new quantum transient stability analysis algorithm achieves outstanding performances both on a quantum simulator and a real IBM quantum device (\emph{ibmq\_boeblingen}, a 20-qubit machine). Therefore, it has the penitential to resolve a grand challenge since the genesis of alternative current power systems.

\section*{Data availability}
The data supporting the findings of this study can be obtained upon reasonable request to the corresponding author.

\bibliographystyle{naturemag-doi}
\bibliography{ref}

\section*{Acknowledgements}
This work was supported in part by the Advanced Grid Modeling Program under the US Department of Energy's Office of Electricity, in part by the US National Science Foundation under Grant Nos. ECCS-2018492 and OIA-2040599, and in part by Stony Brook University's Office of the Vice President for Research through a Quantum Information Science and Technology Seed Grant. We would like to thank the Brookhaven National Laboratory operated IBM-Q Hub. 
\section*{Author contributions statement}

Y.Z and P.Z together devised the idea, concept and methodology, analysed the results and wrote the manuscript.
Y.Z developed the algorithms and conceived the experiments.
P.Z led the research project and supervised the study.

\section*{Competing interests}

The authors declare no competing interests.

\section*{Additional information}

\textbf{Correspondence} and requests for materials should be addressed to P.Z.

\end{document}

%% file: preamble.tex
\usepackage[ruled, linesnumbered]{algorithm2e} %
\usepackage{amsfonts}
\usepackage{amsmath}
\usepackage{amsthm}
\usepackage{amssymb}
\usepackage{array}
\usepackage{bm}
\usepackage{cases}
\usepackage{color}
\usepackage[numbers,sort&compress]{natbib}
\usepackage[font=footnotesize,labelfont=bf]{caption}
\usepackage{enumitem}

\usepackage[overload]{empheq}
\usepackage{graphicx}   
\usepackage{makecell}
\usepackage{mathrsfs}
\usepackage{multirow} 
\usepackage{nameref}

\usepackage{physics}    
\usepackage{soul,color}
\usepackage{subcaption}
\usepackage{url}
\usepackage[all]{xy}

\usepackage{threeparttable}
\usepackage{tabularx}

\usepackage{xspace}
\usepackage{float}

\usepackage{tikz}    
\usepackage{standalone}
\usetikzlibrary{arrows , arrows.meta, calc , fit}
\usetikzlibrary{shapes, chains, fit,backgrounds,shapes,positioning}
\usepackage{marvosym}
\usepackage{circuitikz}
\usetikzlibrary{arrows.meta}

\newcommand{\PreserveBackslash}[1]{\let\temp=\\#1\let\\=\temp}
\newcolumntype{C}[1]{>{\PreserveBackslash\centering}p{#1}}
\newcolumntype{R}[1]{>{\PreserveBackslash\raggedleft}p{#1}}
\newcolumntype{L}[1]{>{\PreserveBackslash\raggedright}p{#1}}

\definecolor{ttextblue}{HTML}{0000FF}
\definecolor{ttextgreen}{HTML}{008000}
\definecolor{ttextred}{HTML}{DC143C}

\newcommand{\QTSA}{\textit{qTSA}}

\usepackage{mathtools}
\usepackage{blkarray, bigstrut}